\documentclass[twocolumn]{aastex6}
\usepackage{amssymb,amsmath,graphicx,natbib,hyperref}
\usepackage{color}
\usepackage{enumerate}
\usepackage{enumitem}

\usepackage[encapsulated]{CJK}
\usepackage[utf8x]{inputenc}
\DeclareGraphicsExtensions{.pdf,.PDF,.png,.PNG,.jpg,.JPG,.jpeg,.JPEG}

\graphicspath{}

\shorttitle{Photometric calibration for BASS}
\shortauthors{Zhou et al.}

\begin{document}
\begin{CJK}{UTF8}{gbsn}

\title{Photometric calibration for the Beijing-Arizona Sky Survey and Mayall z-band Legacy Survey}
\author{Zhimin Zhou\altaffilmark{1}, Xu Zhou\altaffilmark{1}, Hu Zou\altaffilmark{1}, Tianmeng Zhang\altaffilmark{1,2}, Jundan Nie\altaffilmark{1}, Xiyan Peng\altaffilmark{1}, Xiaohui Fan\altaffilmark{3}, Linhua Jiang\altaffilmark{4}, Ian McGreer\altaffilmark{3}, Jinyi Yang\altaffilmark{3}, Arjun Dey\altaffilmark{5}, Jun Ma\altaffilmark{1,2}, Jiali Wang\altaffilmark{1}, Xu Kong\altaffilmark{6}, Qirong Yuan\altaffilmark{7}, Hong Wu\altaffilmark{1,2}, David Schlegel\altaffilmark{8}}

\altaffiliation{$^1$Key Laboratory of Optical Astronomy, National Astronomical Observatories, Chinese Academy of Sciences, Beijing, 100012, China, \email{zmzhou@bao.ac.cn}}
\altaffiliation{$^2$College of Astronomy and Space Sciences, University of Chinese Academy of Sciences, Beijing 100049, China}
\altaffiliation{$^3$ Steward Observatory, University of Arizona, Tucson, AZ 85721}
\altaffiliation{$^4$ Kavli Institute for Astronomy and Astrophysics, Peking University, Beijing 100871, China}
\altaffiliation{$^5$ National Optical Astronomy Observatory, Tucson, AZ 85719}
\altaffiliation{$^6$ Key Laboratory for Research in Galaxies and Cosmology, Department of Astronomy, University of Science and Technology of China, Hefei 230026, China}
\altaffiliation{$^7$ Department of Physics, Nanjing Normal University, WenYuan Road 1, Nanjing 210046, China}
\altaffiliation{$^{8}$ Lawrence Berkeley National Labortatory, Berkeley, CA 94720}

\begin{abstract}
	We present the photometric calibration of the Beijing-Arizona Sky Survey (BASS) and Mayall z-band Legacy Survey (MzLS), which are two of the three wide-field optical legacy imaging surveys to provide the baseline targeting data for the Dark Energy Spectroscopic Instrument (DESI) project. The method of our photometric calibration is subdivided into the external and internal processes. The former utilizes the point-source objects of Pan-STARRS1 survey (PS1) as the reference standards to achieve the zero points of the absolute flux for individual exposures. And then the latter revise the zero points to make them consistent across the survey based on multiple tilings and large offset overlaps. Our process achieves a homogeneous photometric calibration over most of the sky with precision better than 10 mmag for $g$ and $r$ bands, 15 mmag for $z$ band. The accuracy of the calibration is better than 1\%  at the bright end (16-18 mag) over most of the survey area.

\end{abstract}

\section{Introduction}

In the last decades, large-scale imaging observations have become indispensable to the research frontiers in astronomy. Recent and upcoming optical wide-area surveys such as the Sloan Digital Sky Survey \citep[SDSS;][]{York2000}, the Pan-STARRS1 survey \citep[PS1;][]{Chambers2016}, the Dark Energy Survey \citep[DES;][]{DES2016MNRAS}, the Large Synoptic Survey Telescope \citep[LSST;][]{Ivezic2008} offer invaluable opportunities to gain a better unraveling of the creation of elements in stars, understanding of galaxy formation and evolution, and testing of our cosmological model.

The basic information derived from the astronomical imaging surveys is the position and brightness of detected signals, related to astrometry and photometry, respectively. Precision photometry is crucial to a number of current and next-generation science projects, such as the studies of dark energy and other cosmology efforts, the measurement of galaxy clustering on large spatial scales, photometric redshift estimates, and the analysis of microlensing, exoplanet transits and variable sources \citep[e.g.,][]{Nie2010, Scolnic2014, Liu2015}. In the measurement uncertainty, photometric calibration makes an important contribution. Therefore, photometric calibration is necessary to be as accurate and uniform as possible in current surveys.

Photometric calibration is a fundamental concern associated with the imaging survey, and its goal is converting the number of photons recorded in images to the physical flux density emitted from a source in the unit of $\rm erg\ s^{-1}\ cm^{-2}\ Hz^{-1}$ or $\rm erg\ s^{-1}\ cm^{-2}\ \text{\AA}^{-1}$, or in the form of magnitudes. The most widely used astronomical magnitude systems are Vega \citep{Hayes1975}, AB \citep{Oke1983} and ST \citep{Stone1996} systems. The Vega magnitude is standard-based with the zero point defined by the magnitude of Vega being zero in all filters. The AB and ST magnitudes are flux-based with the zero points defined in terms of a reference flux in physical units. 

Accurate photometric calibration is a complicated process, and can be performed with several different approaches. The widely used method is to multiply observe a series of spectrophotometric standard stars to estimate the observation condition such as airmass, transparency, throughput and response of the detector, and then to determine zero points for the science exposures. 
The calibration images are usually exposed immediately following the science observation blocks with similar airmass \citep[e.g.,][]{Brammer2016}, or observed at different airmasses at the beginning and end of each night \citep[e.g.,][]{Drlica2017}, or obtained with an auxiliary telescope like SDSS \citep{Hogg2001}.

The most widely used standard star network is the catalog of Landolt stars \citep{Landolt1992} which contains the Johnson-Kron-Cousins $\it UBVRI$ magnitudes of hundreds of equatorial stars. Besides that, the calibration process also utilizes some extensive catalog of stellar magnitudes in the wide-area sky surveys like SDSS, PS1, the Two Micron All Sky Survey \citep[2MASS;][]{Skrutskie2006}. For example, the Palomar Transient Factory \citep[PTF;][]{Law2009} utilizes SDSS-DR7 PhotoPrimary point sources as photometric standards \citep{Ofek2012}. Besides the external process for the absolute photometric calibration, the internal or relative calibration also should be performed to achieve high photometric uniformity over wide areas of one survey \citep{Padmanabhan2008}. Internal calibration uses the observing strategy of multiple tilings and large offset overlaps to revise the zero points of individual science exposures. 

In addition, other issues also should be considered in the calibration processing like transformations and color term between different photometric systems, correction of star flats, the effect of different types of standards.

In this work, we describe the photometric calibration for the Beijing-Arizona Sky Survey \citep[BASS;][]{Zou2017b} and Mayall z-band Legacy Survey \citep[MzLS;][]{Silva2016}. BASS and MzLS are two wide-field photometric surveys covering the same area of about 5400 deg$^2$ in the northern Galactic cap. The two surveys are performed in parallel. BASS is conducted in $\it g$ and $\it r$ filters, and MZLS is imaged with $\it z$ band. The effective wavelengths are 4776 \AA, 6412 \AA, and 9203 \AA\ for $\it grz $ bands, respectively. The expected 5$\sigma$ magnitude limitings are $\it g=24.0$, $\it r=23.4$, and $\it z=23.0$ for the extragalactic sources after the correction of Galactic extinction. 

BASS utilizes the 2.3 m Bok telescope of the Steward Observatory at Kitt Peak. This telescope has a wide-field camera, 90Prime, equipped with an 2$\times$2 array of active CCDs, each 4K$\times$4K pixels. Each BASS image covers 1.08$\times$1.03 deg$^2$ with a pixel scale of 0\farcs454 pixel$^{-1}$. MzLS is carried out by the MOSAIC-3 camera, a prime focus imaging system on the 4-meter Mayall Telescope at Kitt Peak. The camera is equipped with four 4K$\times$4K pixel CCDs, and images a field of view (FoV) 36\farcm$\times$36\farcm\ with pixels of 15 $\mu$m that subtend 0\farcs26 on a side.

BASS and MzLS, together with the Dark Energy Camera Legacy Survey \citep[DECaLS;][]{Blum2016}, are the three optical legacy imaging surveys, which will provide the baseline targeting data for the Dark Energy Spectroscopic Instrument \citep[DESI;][]{DESI2016a,DESI2016b} survey project. DESI will obtain 30 million galaxy and quasar redshifts spanning over 14000 deg$^2$ to study baryon acoustic oscillations (BAO) and the growth of structure in the universe.

The article is organized as follows. Section \ref{Method} presents the main steps of our photometric calibration pipeline for BASS and MzLS images. In Section \ref{Results} we describe the main outputs of our calibration and provide statistical analysis of the zero points. Section \ref{Discussion} discusses the variation and systematic errors of the zero points. Finally, we summarize our results in Section \ref{Conclusion}.

\begin{figure*}
	\centering
	\includegraphics[width=\hsize]{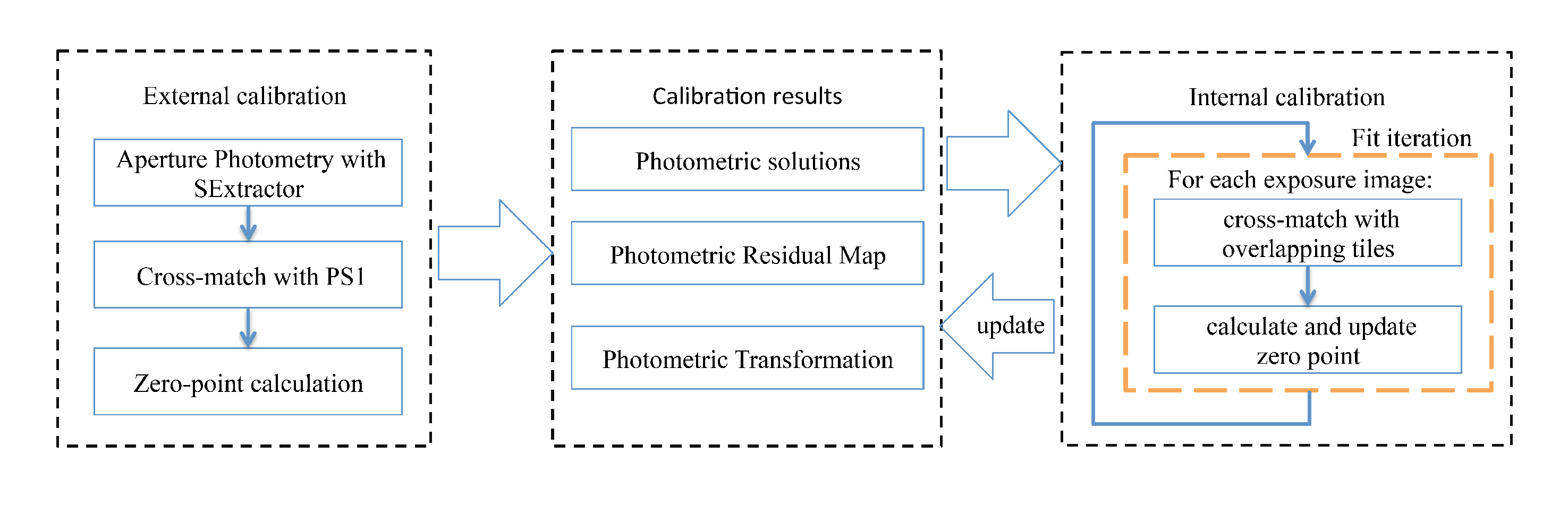}
	\caption{Flowchart of the flux calibration}
	\label{fig.flowchart}
\end{figure*}

\section{Photometric-Calibration Method}
\label{Method}
BASS and MzLS have two main pipelines of data reduction. The first is hosted by the National Optical Astronomy Observatory (NOAO) of United States, includes image processing, calibration, catalog construction and data release which are carried out at the National Energy Research Scientific Computing Center (NERSC) of Lawrence Berkeley National Laboratory (Arjun Dey et al. 2018, in preparation). In this article, we will have no further mention of this pipeline, and only focus on the second pipeline relevant here, as discussed subsequently.

The method of photometric calibration described in this paper is performed in the second pipeline. This pipeline, hosted by the National Astronomical Observatory of China (NAOC), implements image reduction and extracts the source catalogs for BASS and MzLS. The processing includes the corrections of overscan, bias, flat, and crosstalk, astrometric and photometric calibration, source extraction and photometry. This pipeline is described in \citet{Zou2017a, Zou2017c}, Tianmeng Zhang et al. (2018, in preparation) and Xiyan Peng et al. (2018, in preparation).

As shown schematically in Figure \ref{fig.flowchart}, our photometric-calibration process comprise two main steps, the external (or absolute) calibration and the internal (or relative) calibration. The former achieves the photometric zero point on the AB magnitude system for each image of each night, filter, and CCD using the PS1 point-source catalog as an external reference. The latter uses multiple tilings and large offset overlaps to revise the zero points of individual science exposures, and obtains a uniform calibration across the survey.

\subsection{External calibration}
The input data of the photometric calibration is the science images after corrections of imaging reduction and astrometric calibration. The two cameras of our surveys both have four CCDs and each CCD is read out by four amplifiers. In the imaging process, each exposure file is split into four smaller FITS images corresponding to four CCDs. On the individual single-epoch CCD sub-images, we use SExtractor software \citep{Bertin1996} to detect the sources and perform the aperture photometry independently, and derive the catalog of instrument magnitude with the preset internal zero point of 0.0.

A large aperture radius of 13 pixels is used in all the calibration measurements. This radius is corresponding to 5\farcs9 for BASS and 3\farcs2 for MzLS, about 3-4 times of the typical seeing for both surveys. This aperture is large enough to be referred as an ``infinite'' aperture, and to neglect the aperture corrections and the spatial variation of the PSF on the chip. Thus, it allows us to estimate the total flux of the bright and isolated point sources used in our calibration, yet the background will not contribute much to the errors. 

From the photometric catalog obtained above, the sources with photometric rms$<$0.1 mag and SExtractor internal flags $<$2 are selected to be the calibration objects. The latter criterion has removed the objects that are deblended, saturated, or have other processing problems. Then the calibration sources are cross-matched with PS1 point-source catalogs with the cross-matching radius of 2\farcs0. For the PS1 catalog, the PSF magnitude is used in the measurement of zero points, and the reference stars are chosen by requiring that:
\begin{enumerate}[itemsep=0pt,parsep=0pt]
\item The rms of the magnitude in the corresponding filter is less than 0.1 mag. 
\item The star was observed and detected at least once in each of the $\it g$, $\it r$, $\it i$, and $\it z$ bands. 
\item The stars should be in the color range of 0.4 $<$ ($\it g$-$\it i$) $<$ 2.7.
\end{enumerate}

The first two criteria select the PS1 objects with good photometric measurements, and the last one selects the main-sequence stars with 0.4 $<$ ($\it g$-$\it i$) $<$ 2.7 which are used to fit the color transformation from PS1 filters to those in our instruments. Although the filters of our surveys are fairly close to the filters in PS1, there are still tiny differences between them, and the correction of color term is needed to convert from PS1 filter wavebands to our photometric system (See Section \ref{colorterm}).  

Based on the cross-matched objects, the difference between PS1 magnitude $MAG_{PS1}$ and our instrument magnitude $MAG_{inst}$ is compared, and objects with bad photometry are rejected with a 3$\sigma$-clipping criterion using the local median and standard deviation. Then the zero point is defined by the inverse variance weighted average of the differences: $ZP=MAG_{PS1}-MAG_{inst}$. Note that the airmass correction has been folded into our zero point, there is no extra airmass correction included in the zero point analysis.

\begin{figure}
	\centering
	\includegraphics[width=\columnwidth]{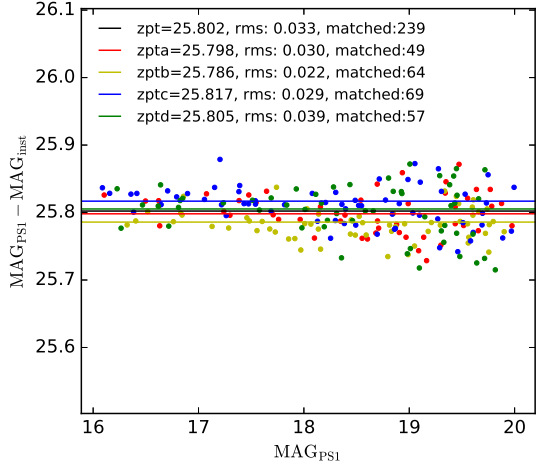}
	\caption{The measurement of zero point for a CCD-image (p7833r0123\_1). The zero points for the whole CCD image (zpt) and four amplifiers (zpta $-$ zptd) are marked by the horizontal lines, and they are calculated individually using the photometric error-weighted average of the magnitude differences between our instruments and PS1. Their uncertainties and the numbers of stars used in the measurements are also marked.}
	\label{fig.zpeg}
\end{figure}

Figure~\ref{fig.zpeg} shows an example of the measurement of the zero point for one CCD-image. There are four readout amplifiers for one CCD, so the zero points for the whole CCD and each amplifier are all calculated, although only the former value is adopted in the internal calibration and the produce of the final photometric catalog. Some exposures may fail to derive their zero points due to a lack of reliable photometric reference stars or beyond the sky region of PS1. For the latter, we calculate a rough temporary zero point using UCAC4 \citep{Zacharias2013} as the reference catalog, and revise it in the internal calibration.

\begin{figure}
	\centering
	\includegraphics[width=\columnwidth]{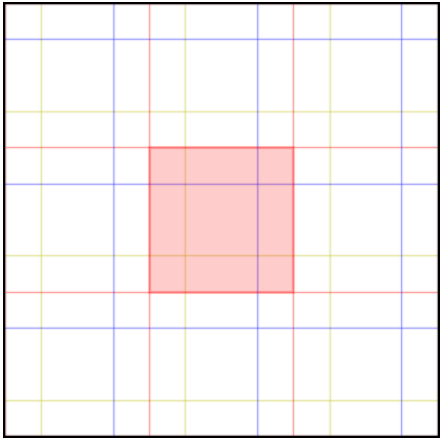}
	\caption{Example of the observing tiling in the sky. Three passes are shown with different colors. Based on the tiling strategy, a single exposure image such as the central one (masked by red color) is overlapped with several neighboring tiles. The internal calibration is performed using the objects in the overlapping regions.}
	\label{fig.tiling}
\end{figure}

\subsection{Internal calibration}

BASS and MzLS adopt a three-pass observing strategy, in which each tile with the size of FoV is observed three times with the dither of about 1/4 FoV each time (Figure \ref{fig.tiling}). Thus, most of the survey area is exposed three times, along with dense overlapping regions between exposures. Multiple tilings and large offset overlaps allow us to establish a consistent photometric calibration across the entire survey region using our algorithm of the internal calibration. In addition, images failed in the external process can also be calibrated here.

This algorithm utilizes the photometric catalogs and zero points obtained in the external calibration, and does not need to consider the filter transformation. For each image, we extract the stellar sources with good photometric measurements utilizing the following criteria: photometric error $<$ 0.1 mag, SExtractor internal flag $<$ 2, ellipticity $<$ 0.2, CLASS\_STAR $>$ 0.75. These sources are corrected by the photometric residual map of each CCD and filter (See Section \ref{photometric residual}), and then cross-matched with those in the overlapping images. Next, a revised zero point for the image is calculated and updated immediately with the calibration candidates in the overlapping images as the reference. This procedure is repeated iteratively until up to the maximum times required (20 times by default) or the criteria that the changes of zero points are less than 0.001 mag for all of the images. In each iteration, all images are calibrated in a random order to avoid error propagation, and the outliers are rejected with sigma clipping of 3$\sigma$.

\begin{figure*}
	\centering
	\includegraphics[width=0.6\hsize]{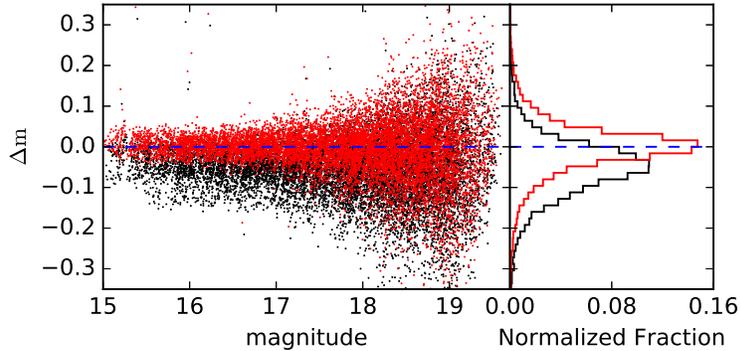}
	\caption{Comparison of the magnitudes of the objects in an image (p7817g0042\_1) with its overlaps. In the left panel, black and red points are the results before and after the correction of internal calibration, their corresponding histogram distributions are shown in the right panel with the black and red lines, respectively.}
	\label{fig.intcali_eg}
\end{figure*}

Figure~\ref{fig.intcali_eg} shows an example of the effect of internal calibration, it compares the object magnitudes in one image (p7817g0042\_1) with those of overlapping exposures before and after the internal calibration. p7817g0042\_1 is beyond the PS1 region, and is calibrated with UCAC4 in the external calibration. We can find that the offset and dispersion obviously become smaller after the internal calibration.

\section{Calibration Results}
\label{Results}

The parameters produced from our calibration pipeline are stored in the headers of the image FITS files. Table~1 summarizes the header keywords related to the photometric solutions. It mainly includes the zero points from the external (CCDZPT) and internal (INT\_ZPT) calibrations, and the related information like calibration scatters, numbers of objects used in the process. The keyword INT\_ZPT is used for calibrating the instrumental magnitudes in the produce of the final stacked images and photometric measurements. In addition, some information about the observation are also added in the header, such as the seeing, limiting magnitude, the color term transformed from PS1 to our system.

\begin{deluxetable*}{ll}
\tablewidth{0pt}
\tablecaption{The keywords of photometric calibration in the header of FITS images}
\tablehead{\colhead{Keyword} & \colhead{Description} }
\label{table:header_key}
\startdata
CALI\_REF & the reference database for calibration       \\   
CCDZPT    & zero point for CCD from external calibration \\                      
CCDZPTA   & zero point for CCD ampA [1:2016,1:2048]      \\   
CCDZPTB   & zero point for CCD ampB [2017:4032,1:2048]   \\ 
CCDZPTC   & zero point for CCD ampC [1:2016, 2049,4096]  \\
CCDZPTD   & zero point for CCD ampD [2017:4032,2049:4096]\\   
CCDPHRMS  & zpt rms of the matched objects in CCD        \\
PHRMSA    & zpt rms of the matched objects in CCD ampA   \\
PHRMSB    & zpt rms of the matched objects in CCD ampB   \\
PHRMSC    & zpt rms of the matched objects in CCD ampC   \\
PHRMSD    & zpt rms of the matched objects in CCD ampD   \\
APER\_R   & photometric-aperture radius (in pixels)      \\                               
SEEING    & seeing in arcsec                             \\
CCDNSTAR  & total number of stars detected on a CCD      \\
NMATCH    & total number of matched stars in 2 arcsec    \\
NMATCHA   & number of matched stars in CCD ampA          \\
NMATCHB   & number of matched stars in CCD ampB          \\
NMATCHC   & number of matched stars in CCD ampC          \\
NMATCHD   & number of matched stars in CCD ampD          \\
MDNCOL    & median (g-i) color of matched stars in CCD   \\
COLT\_PAR & parameters in $[gi^3,gi^2,gi^1,gi^0]$        \\
MLIM      & magnitude limiting of 5-sigma galaxy detection\\             
INT\_ZPT  & zero point from internal calibration          \\
INT\_STD  & sigma for zpt from internal calibration       \\
INT\_NUM  & number of stars used in internal calibration  \\
\enddata
\end{deluxetable*}

\begin{figure*}
	\centering
	\includegraphics[width=0.6\hsize]{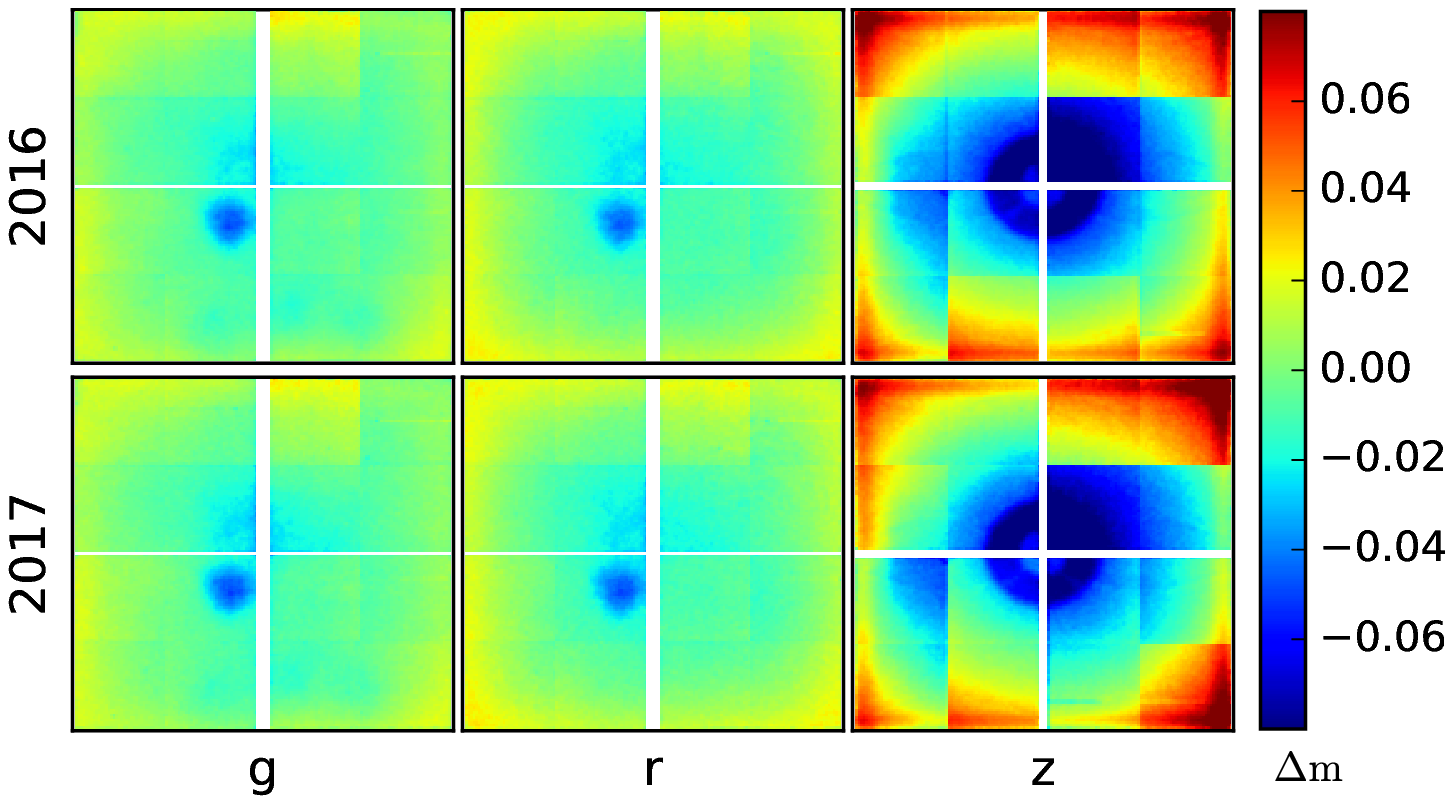}
	\caption{Photometric residual maps for the $\it grz$ filters. For each band, the maps of 2016 and 2017 seasons are shown. The maps clearly show the jumps between different amplifiers and CCDs, the patterns and signs of scattered light, especially for $z$ band.}
	\label{fig.residualmap}
\end{figure*}
 
\subsection{Photometric Residual Map}
\label{photometric residual}

Based on the calibrated catalogs, the photometric residual maps are established for each CCD of each filter. For each image, we select matches between the PS1 reference stars and observed stars with good photometric measurements in the image, and measure the differences between the reference magnitudes and observed calibrated magnitudes. Then the maps of the differences are combined and averaged using thousands of the science exposures. Because the observing strategy of our surveys are multiple pointings with large dithers, objects can be situated on any position of CCD images in principle. Thus, this step can construct the two-dimensional residual maps (in magnitudes) with the same pixels to our reduced images.

The produced photometric residual maps, also named photometric flat or star flat, are different from the standard flat fields used in the imaging process. The latter is usually obtained nightly from a uniformly illuminated field (such as dome, twilight and sky), and is mainly to correct the varying efficiency of the individual pixels for each observing night. Photometric residual maps are mainly caused by geometric distortion from variations across the focal plane, low-frequency structure from the imperfect illumination of the flat-field screen, vignetting and scattered light compensation \citep{Tucker2007,Tucker2018}. In addition, they can be applied directly to the measured photometry.

Our residual maps have proven to vary slightly due to the maintenance of the telescope every summer. We therefore generate residual maps for each observing season with every August as the boundary. For example, the 2017 season is from September 2016 to July 2017. These maps are shown in Figure \ref{fig.residualmap}. For the BASS system, the standard deviation is $\sim$ 10 mmag in the maps of $g$ and $r$ filters, the mean absolute difference is less than 5 mmag between the maps of different seasons. For MzLS $z$ band, the standard deviation is $\sim$ 30 mmag for each CCD, and the seasonal variation is 5-10 mmag for each CCD from 2016 to 2017.
The maps clearly show the jumps between different amplifiers and CCDs, the patterns and signs of scattered light, suggesting that a photometric residual correction would be valuable. Especially for $z$ band, the bright ring present in the residual fields indicates that the center of the focus plane is too faint and needed correction. 

\begin{figure*}
	\centering
	\includegraphics[width=0.6\hsize]{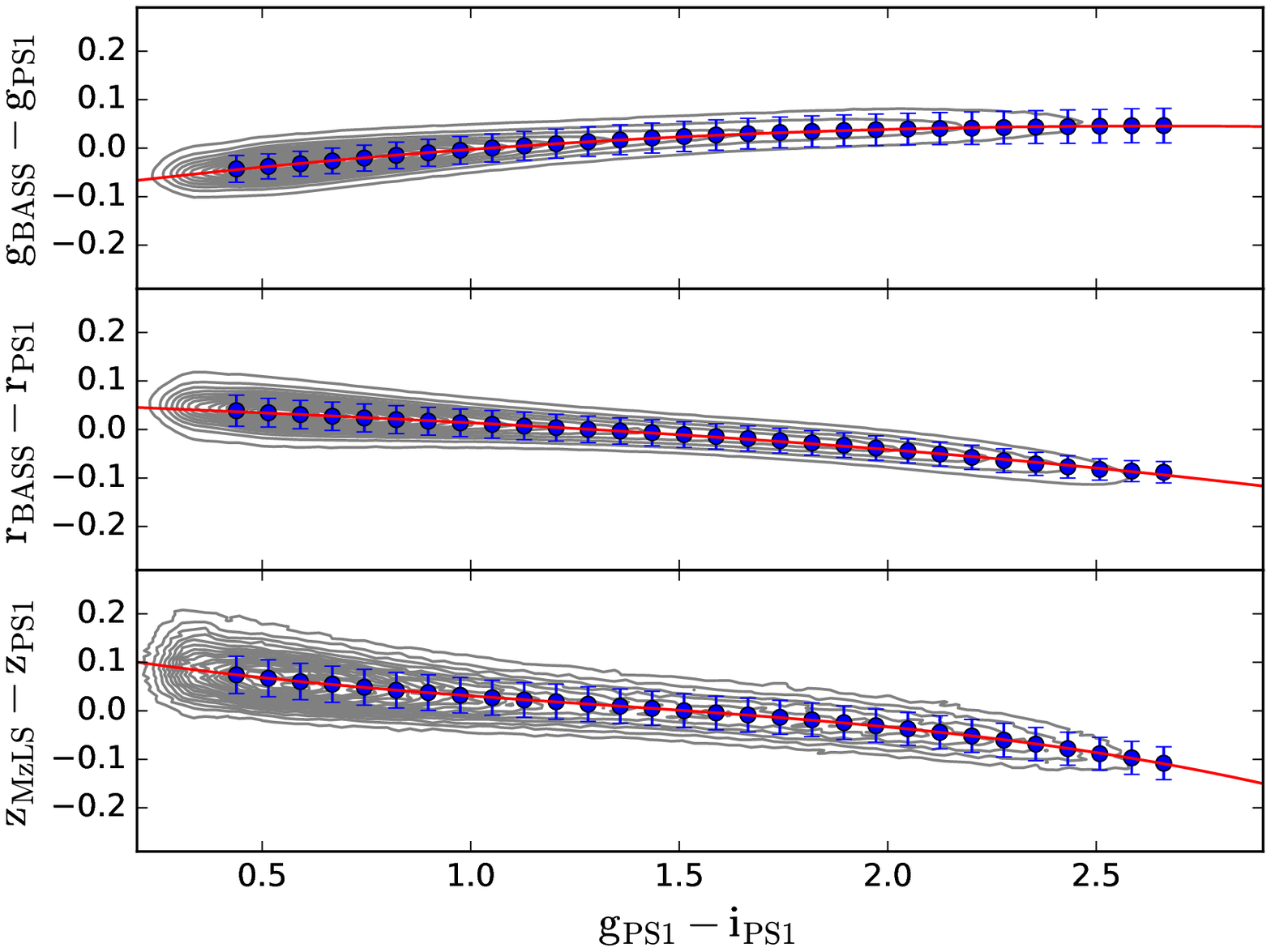}
	\includegraphics[width=0.6\hsize]{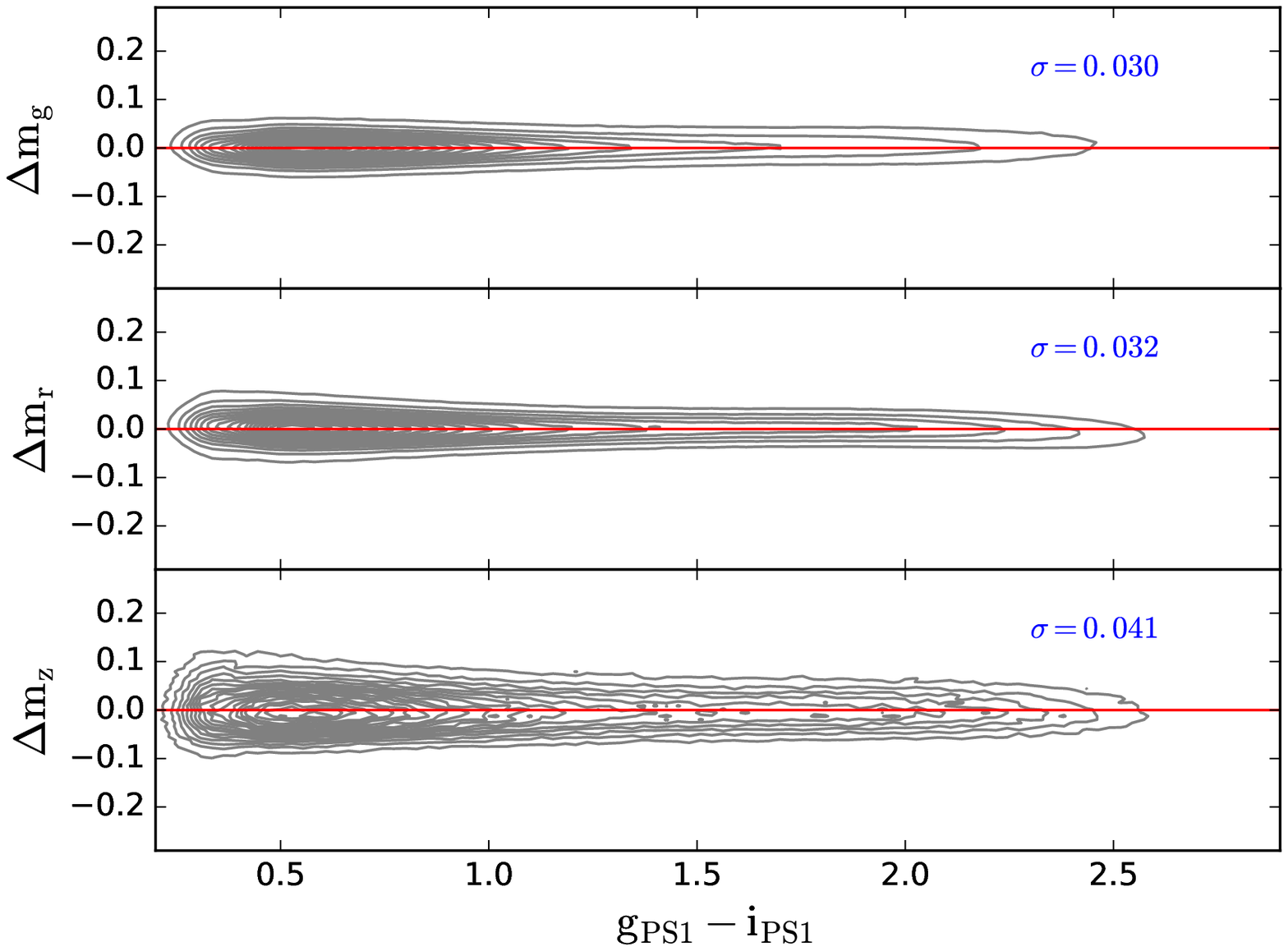}  
	\caption{Color transformations from PS1 system to BASS/MzLS system. The top three panels are for BASS $g$, $r$ and MzLS $z$ band, respectively. The contours show the distributions of the  magnitude differences between PS1 and BASS/MzLS. The transformations are as a function of $g - i$. The median values and deviations in the color bins are shown as blue points with error bars. The third-order polynomial fittings are shown as red lines. In the bottom, the panels show the residuals of the fittings for the three bands along with their standard deviations labeled. }
	\label{fig_colorterm}
\end{figure*}

\subsection{Photometric Transformation}
\label{colorterm}

In general, there are mismatches between the effective bandpasses of our filter system and those of the PS1 system, although they are very close to each other. Thus, color terms are needed to convert from the PS1 reference filter wavebands to the instrumental wavebands in our surveys.

We select the main-sequence stars with $0.4 < \it{g} - \it{i} < 2.7$ mag from PS1 reference catalog, compare the magnitude difference between our survey and PS1 as a function of color index ($\it g - i$), and then derive the fit using a third-order polynomial (Figure \ref{fig_colorterm}). In order to ensure no systematic offsets between the magnitudes of PS1 and our surveys, a constant term is added to make the mean values of the color terms equal to zero. Therefore, the $g$-, $r$- and $z$-band magnitudes in PanSTARRS1 are transformed into the BASS and MzLS frame using the following color equations:
\begin{equation}
\begin{aligned}
(g-i) = g_{\rm PS1}-i_{\rm PS1}\\
g_{\rm BASS} = g_{\rm PS1} - 0.08826 + 0.10575(g-i)\\
 - 0.02543(g-i)^2 + 0.00226(g-i)^3 \\
r_{\rm BASS} = r_{\rm PS1} + 0.07371 - 0.07650(g-i)\\
 + 0.02809(g-i)^2 - 0.00967(g-i)^3 \\ 
z_{\rm MzLS} = z_{\rm PS1} + 0.10164 - 0.08774(g-i)\\
 + 0.03041(g-i)^2 - 0.00947(g-i)^3 
\end{aligned}
\end{equation}


\section{Discussion}
\label{Discussion}

We have computed the photometric calibration for the observations of BASS/MzLS, derived zero points for each image, and solved the photometric flat and color transformation for each filter. In this section, we make statistic and analyze these results, check the accuracy and consistency of the calibration solution, and test the stability of the survey systems.

\subsection{Statistics of Zero Points}
\begin{figure*}
	\centering
	\includegraphics[width=0.8\hsize]{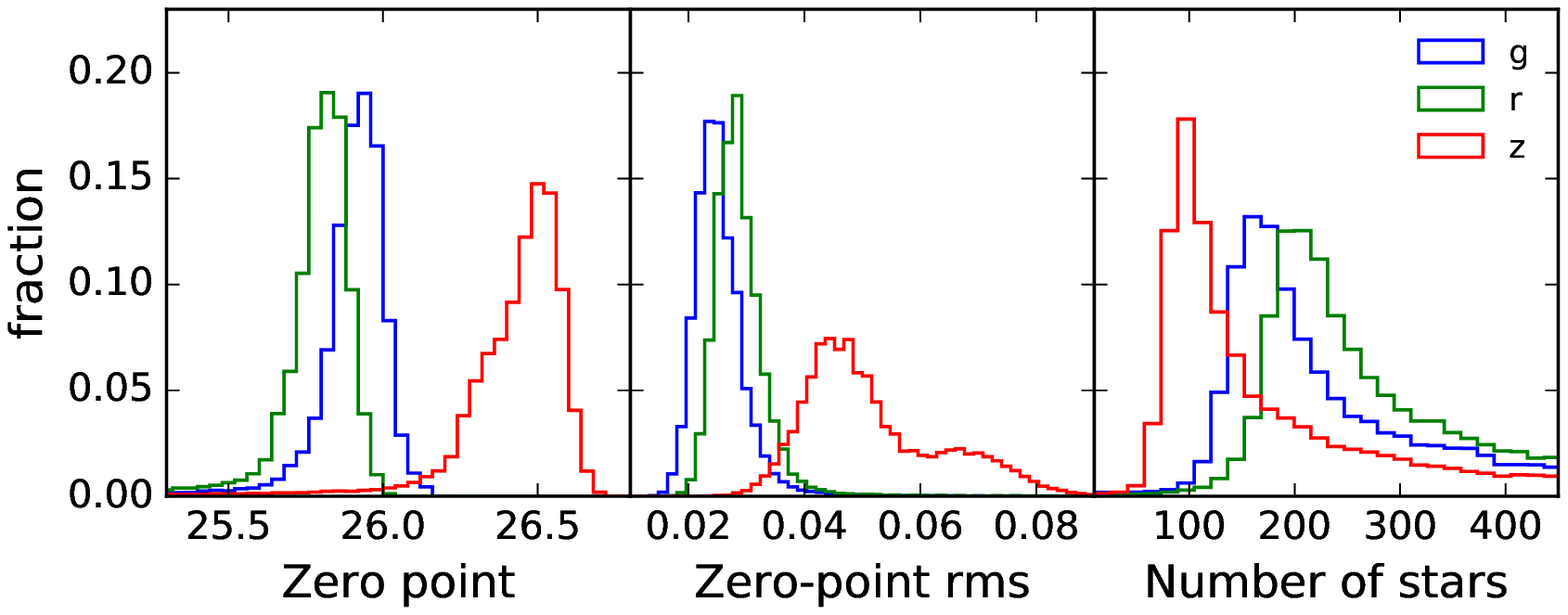}
	\caption{Histograms of the zero point, rms of the zero point, and the number of reference stars used in the measurement. In each panel, $grz$ bands are marked with blue, green, and red lines, respectively.}
	\label{fig_zpt1}
\end{figure*}

\begin{figure*}
	\centering
	\includegraphics[width=\hsize]{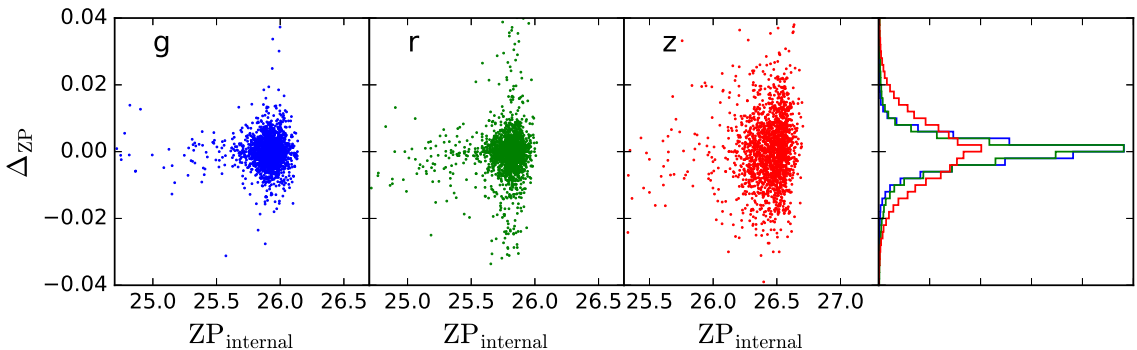}
	\caption{Comparisons between the zero points before and after the correction of internal calibration for the filters $grz$. The left three panels give the changes of the zero points from external calibration relative to those derived from the internal correction. The distribution of the differences in magnitudes is shown on the right histogram plot, and the $g$, $r$ and $z$ band are shown with the blue, green, and red colors, respectively.}
	\label{fig_zpt2}
\end{figure*}

\begin{figure*}
	\centering
	\includegraphics[width=\columnwidth]{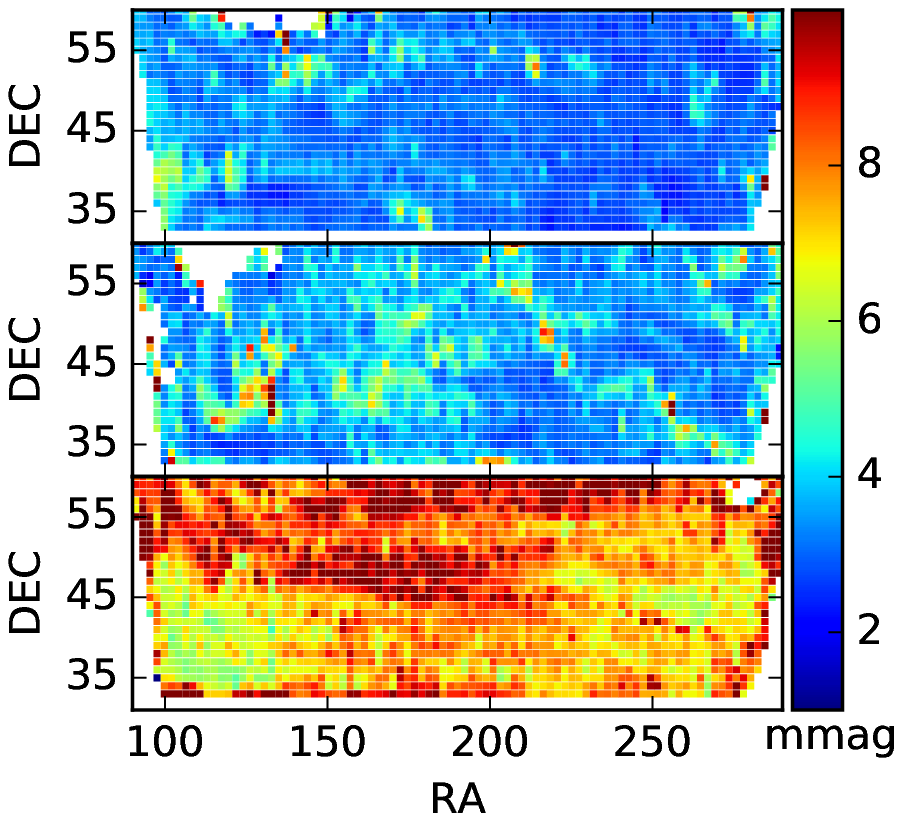}
	\includegraphics[width=\columnwidth]{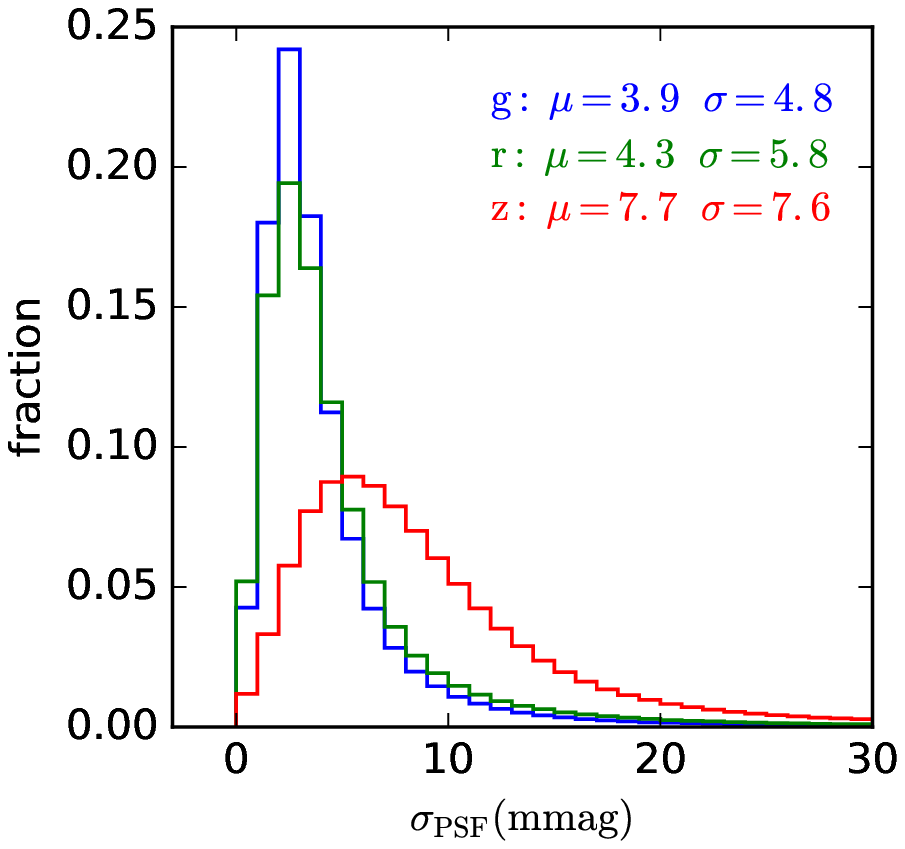}
	\caption{Maps (left panel) and histograms (right panel) of the dispersions for repeated measurements on the PSF magnitude. The dispersion is computed based on the stars with 16-18 mag and at least 3 exposures in a given band. In the left panel, three maps show the dispersions for $g$, $r$ and $z$ band from top to bottom, respectively. The x-axes give the right ascension and the y-axes give declination, both in degrees.	The three histograms in the right panel correspond to the distributions over the survey for the three bands $grz$. The median $\mu$ and standard deviation $\sigma$ of the magnitude rms $\sigma_{PSF}$ are labeled for each filter.}
	\label{fig_consis}
\end{figure*}

Figures \ref{fig_zpt1} characterizes the statistical distribution of the results of external calibration, including the zero point, its measured error, and the number of stars used in the measurement. The median zero point is 25.92, 25.80, and 26.46 in magnitude for the filters $grz$, respectively. 
The number of objects used for the calibration of each CCD image varies greatly depending on the sky coordinates of the exposures, clearly increasing when the position near the Galactic plane. In addition, it varies slightly with the FOV area, the data quality, and the observational condition. In general, the median number of available stars is around 180 for $g$ band, 220 for $r$ band, and 100 for $z$ band, corresponding to 700-1100 per square degree. The corresponding measured error of the zero point is about 0.02, 0.02, 0.05 mag for the three filters. Considering the number of available stars in each image, the precision (here the standard deviation of the mean is used) is about 0.001, 0.002, and 0.004 mag for $g$, $r$ and $z$, respectively.

The internal calibration provides a consistency correction for all the exposures over the entire survey. Figure \ref{fig_zpt2} verifies the results of the internal calibration by comparing the zero points generated by the external measurements with those after the internal correction. Zero points derived from two processes are consistent. Unsurprisingly, the mean offset is near zero as expected. The scatter of their differences is about 4 mmag, 6 mmag, and 10 mmag for $grz$ bands, respectively. These scatters also reflect the accuracy of our external calibration.

\begin{figure*}
	\centering
	\includegraphics[width=\columnwidth]{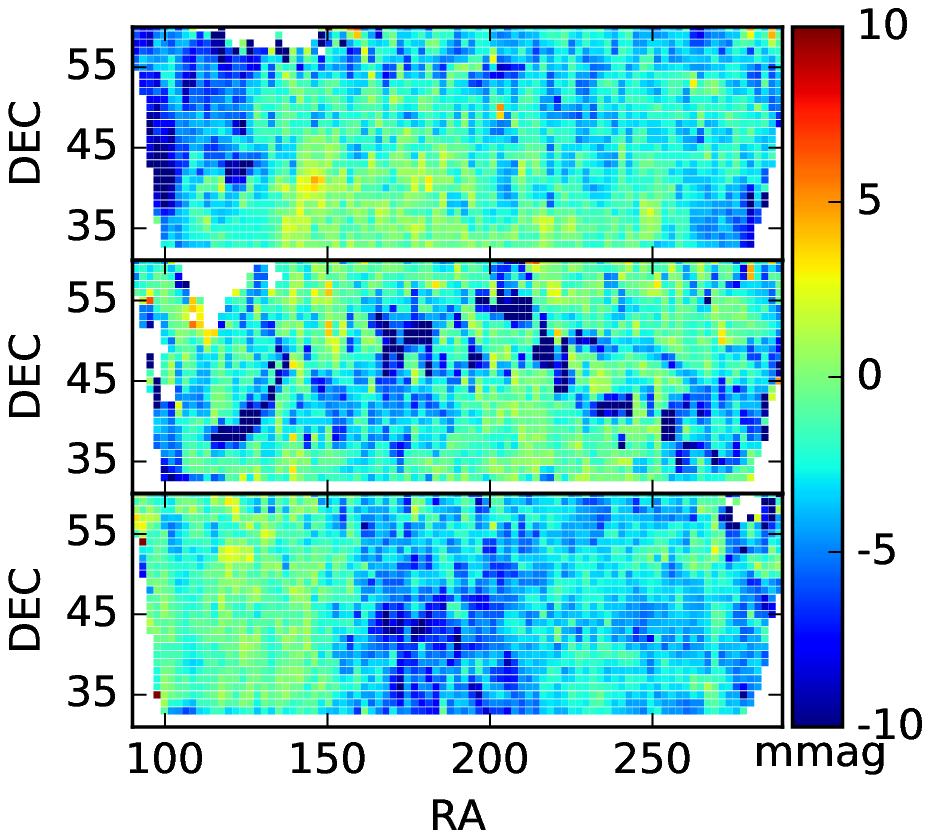}
	\includegraphics[width=\columnwidth]{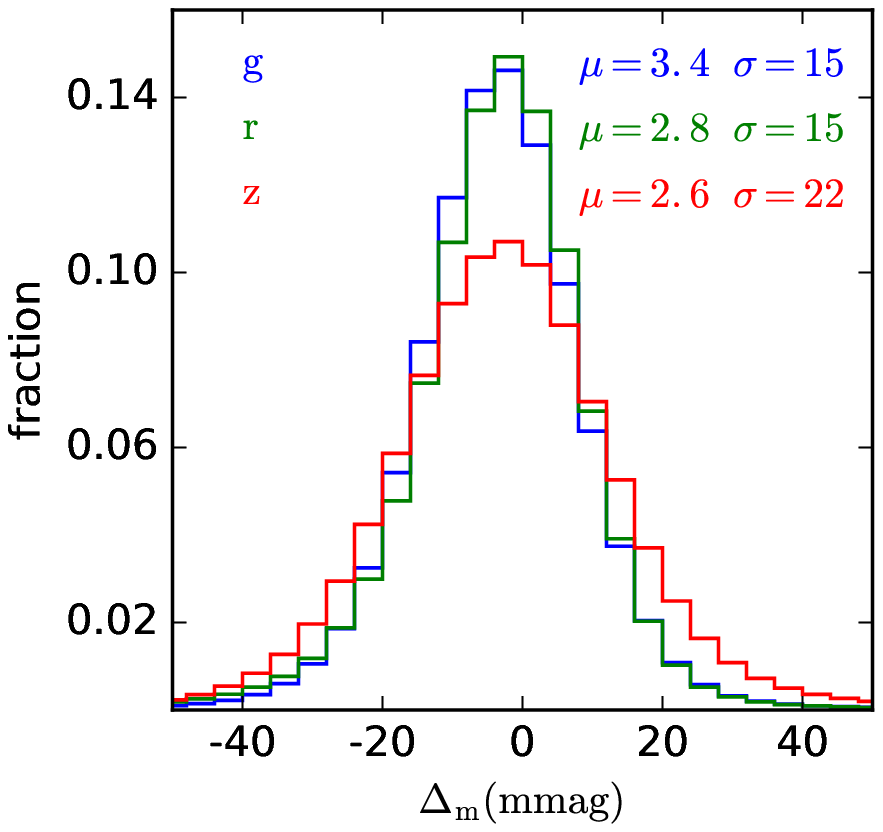}
	\caption{Comparison of PSF magnitude from our survey and PS1. The left panel shows the maps of the median difference between the PSF magnitudes of stars and the color-transformed PS1 magnitude of the same stars in the filters $grz$ (from top to bottom). The histograms in the right panel show the overall distribution of the differences over the survey. The median $\mu$ and standard deviation $\sigma$ of the magnitude offset are labeled for each filter.}
	\label{fig_vsps1}
\end{figure*}

\subsection{Analysis of the Consistency}

In order to evaluate the precision of our calibration, we check the internal consistency based on the PSF photometry from BASS Data Release 2 \citep{Zou2017c}. We select stars over the range 16-18 mag and with at least 3 exposures in a given filter, derive the PSF magnitude rms of multiple observations. Figure \ref{fig_consis} shows the distribution of the magnitude rms for $grz$ filters in our survey. The median rms is only about 3.9, 4.3, 7.7 mmag with rms scatter of 4.8, 5.8 and 7.6 mmag in $grz$ filters. There is a larger rms distribution for $z$ band mainly due to the brighter magnitude limiting and pattern noise of $z$-band images. For a constant source, the magnitude rms mainly combines the uncertainties related to the photometric error and zero point of our calibration. Thus, the actual consistency or precision of our zero points should be better than 10 mmag in $g$, $r$, and $z$.

To look for possible spatial structure in the calibration, Figure \ref{fig_consis} also shows maps of the rms of PSF magnitudes over the survey. In general, the rms distribution are homogeneous over most of the sky, only a few isolated regions have slightly increased scatter than others. As we find in the histogram, $z$-band map has a larger rms distribution over the footprint than $g$ and $r$ band.


\begin{figure*}
	\centering
	\includegraphics[width=\hsize]{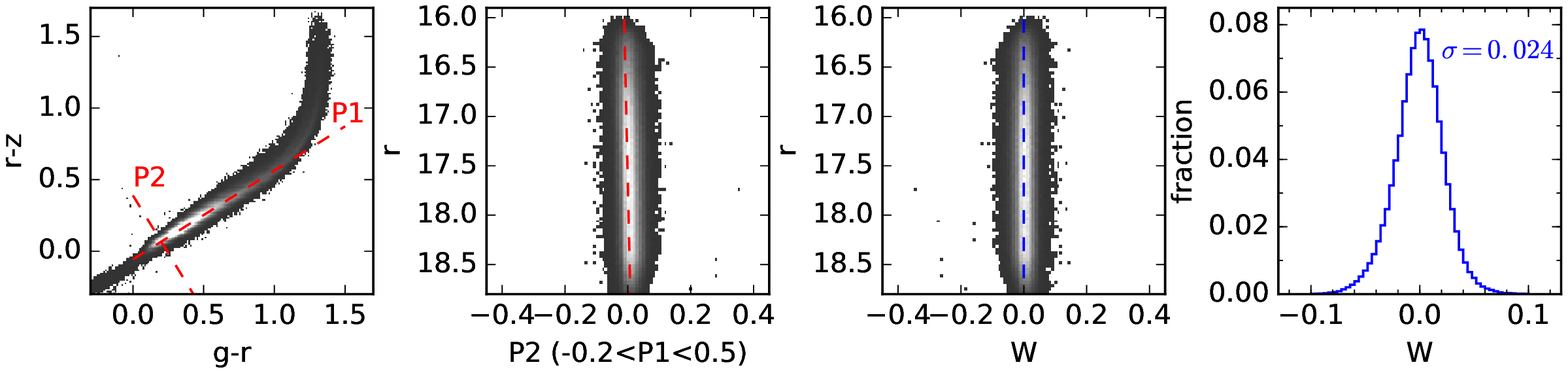}
	\caption{Demonstration of principal color fitting in the color-color diagram. The first panel on the left displays the $g-r$ vs. $r-z$ diagram. The red lines show the linear fits to parallel (P1) and perpendicular (P2) to the blue part of the stellar locus. The second panel shows the P2 color as a function of $r$ magnitude along with the linear fit (the red line). The third panel shows $W$, the corrected version of principal color P2, as a function of $r$ magnitude.The vertical blue line marks $W=0$. The last panel shows the histogram distribution of $W$, and its rms scatter is 0.024.}
	\label{fig_color_locus1}
\end{figure*}

\begin{figure}
	\centering
	\includegraphics[width=\columnwidth]{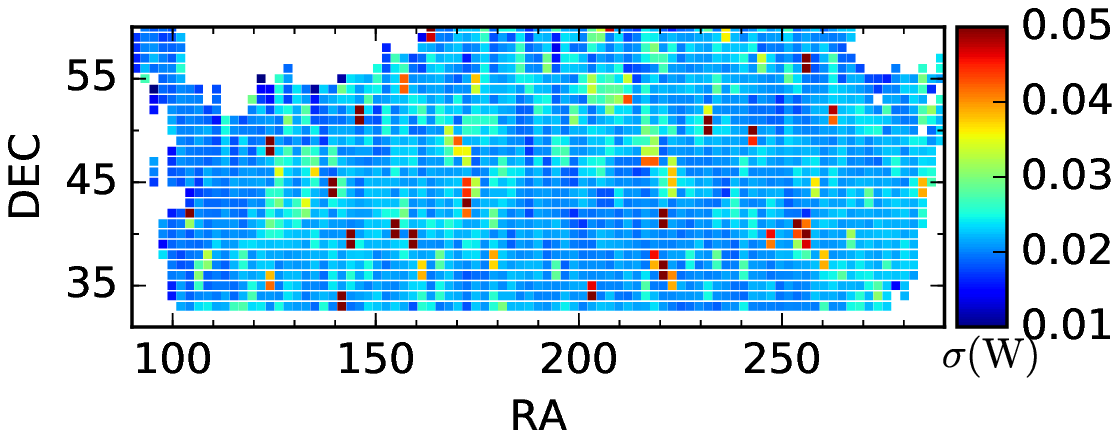}
	\caption{Map of the rms scatter of the stellar locus W over the sky.}
	\label{fig_color_locus2}
\end{figure}

\subsection{Analysis of the Accuracy}

To verify the accuracy of photometric calibration, we cross match the stars selected in the above consistency test with PS1 reference catalog, and then calculate the differences between the calibrated PSF magnitude of objects and their color-transformed PS1 magnitude. The results of the comparison are shown in Figure \ref{fig_vsps1}. The histograms show the overall distribution of the differences in our survey. The median offset between two sets of measurements is very close to zero (2-3 mmag) in all three bands, the dispersion is about 15 mmag for $gr$, and 22 mmag for $z$. 

The maps in Figure \ref{fig_vsps1} further show the distribution of mean differences for these magnitudes to examine the calibration accuracy over the sky. Although there are a small number of regions with large magnitude offset, most of the footprints have the offset less than 10 mmag. We further find that the regions with bad accuracy generally have bad consistency, suggesting non-photometric observation of these regions. The quality of these regions may be better as the number of overlapping observations increases.

\subsection{Consistency of Stellar Locus}

We additionally check the calibration quality with the stellar locus analysis described by \citet{Ivezic2004,Ivezic2007}. The stellar locus is a tight one-dimensional locus of main-sequence stars in color-color diagrams. Its intrinsic width is quite small ($\sim$ 0.01 mag), comparable to the calibration error. Most of stars detected by our surveys are main-sequence stars, the scatter of the stellar locus is dominated by the errors in photometric zeropoint calibration, so it can be used as a diagnostic of calibration quality.

Following the methodology in \citet{Ivezic2004}, we construct the principal color which tracks the direction perpendicular to the stellar locus. Because there are $g$, $r$, $z$ bands in our surveys, the principal color $W$ is defined based on the blue part of the locus in the $g-r$ vs. $r-z$ diagram (Figure \ref{fig_color_locus1}). We select only bright stars with 16-19 mag and good measurements in all three filters, and use the linear fits to determine principal colors (P1, P2) parallel and perpendicular to the locus in the color-color diagram. While there is a small trend of the P2 color as a function of $r$ magnitude, so we again derive a linear fit and obtain the corrected version of P2 as the final principal color: $W = 0.527g - 1.370r + 0.850z - 0.168$.

The histogram on the right panel of Figure \ref{fig_color_locus1} shows the distribution of the principal color $W$. The distribution rms width is 0.024, indicating the accuracy of calibration to $\sim$ 1-2\%, consistent with our above estimates. Figure \ref{fig_color_locus2} shows  the variation of the principle color over of survey. Over most of the survey footprint, the calibration errors are found to be uniform, while a small number of large photometric outliers exist.



\section{Conclusion}
\label{Conclusion}
We have described the photometric calibration of BASS and MzLS. BASS and MzLS are two of the three wide-field optical legacy imaging surveys, which will provide the baseline targeting data for DESI project. The current process of our photometric calibration is divided into two main steps, the external and internal calibration. The external calibration process utilizes PS1 point-source objects as photometric standards, and achieves the photometric zero point on the AB magnitude system for each image of each night, filter, and CCD. Based on multiple tilings and large offset overlaps of the survey, the internal calibration process uses the calibrated stars in the overlapping images to revise the zero points of individual science exposures, and obtains a uniform calibration across the survey.

As the products of the calibration, the parameters about zero point are stored in the keywords of the image FITS header. In addition, color transformations from PS1 system to BASS/MzLS are derived, and the photometric residual map is obtained for each CCD and each filter.

We have checked the accuracy and consistency of our calibration solution. In general, the internal consistency over the sky is better than 10 mmag for $g$ and $r$ filter, better than 15 mmag for $z$ filter. At the bright end (16-18 mag), the accuracy of the absolute calibration is better than 1\% relative to PS1 over the survey area. The check with the stellar locus analysis also confirms the results. Although there are a small number of regions with slightly large uncertainness, their calibration quality will be better as the number of overlapping observations increases.
 
\acknowledgements
\label{sec:acknow}
We thank the anonymous referee for numerous valuable suggestions and comments.
The BASS is a collaborative program between the National Astronomical Observatories of Chinese Academy of Science and Steward Observatory of the University of Arizona. It is a key project of the Telescope Access Program (TAP), which has been funded by the National Astronomical Observatories of China, the Chinese Academy of Sciences (the Strategic Priority Research Program ''The Emergence of Cosmological Structures`` Grant No. XDB09000000), and the Special Fund for Astronomy from the Ministry of Finance. The BASS is also supported by the External Cooperation Program of Chinese Academy of Sciences (Grant No. 114A11KYSB20160057) and Chinese National Natural Science Foundation (Grant No. 11433005). The BASS data release is based on the Chinese Virtual Observatory (China-VO).

This work was supported by the Chinese National Natural Science Foundation grands No. 11673027, 11373035, and 11603034 by National Key R\&D Program of China (Grant No. 2017YFA0402704) and by the National Basic Research Program of China (973 Program), No. 2014CB845704, 2013CB834902, 2014CB845702, and 2015CB857004.

The Pan-STARRS1 Surveys (PS1) and the PS1 public science archive have been made possible through contributions by the Institute for Astronomy, the University of Hawaii, the Pan-STARRS Project Office, the Max-Planck Society and its participating institutes, the Max Planck Institute for Astronomy, Heidelberg and the Max Planck Institute for Extraterrestrial Physics, Garching, The Johns Hopkins University, Durham University, the University of Edinburgh, the Queen’s University Belfast, the Harvard-Smithsonian Center for Astrophysics, the Las Cumbres Observatory Global Telescope Network Incorporated, the National Central University of Taiwan, the Space Telescope Science Institute, the National Aeronautics and Space Administration under Grant No. NNX08AR22G issued through the Planetary Science Division of the NASA Science Mission Directorate, the National Science Foundation Grant No. AST-1238877, the University of Maryland, Eotvos Lorand University (ELTE), the Los Alamos National Laboratory, and the Gordon and Betty Moore Foundation.

\bibliographystyle{apj}
\bibliography{}

\end{CJK}
\end{document}